\documentclass[a4paper,fleqn,twoside]{article}
\usepackage{espcrc2}
\usepackage{graphics}
\usepackage{mathbbol}

\title{First Simulation Results for the Photon in a Non-Commutative Space}
\author{W. Bietenholz \address[HU]{Institut f\"ur Physik, Humboldt 
Universit\"at zu Berlin, Newtonstr.\ 15, 12489 Berlin, Germany},
F. Hofheinz \addressmark[HU],
J. Nishimura \address[Japan1]{High Energy Accelerator Research 
Organization (KEK), 1-1 Oho, Tsukuba 305-0801, Japan},
Y. Susaki \addressmark[Japan1]\addressmark[Japan2]
\address[Japan2]{Institute of Physics, 
University of Tsukuba, Tsukuba, Ibaraki 305-8571, Japan} and
J. Volkholz \addressmark[HU]}

\begin{document}

\begin{abstract}
We present preliminary simulation results for QED in a non-commutative
4d space-time, which is discretized to a fuzzy lattice. Its numerical
treatment becomes feasible after its mapping onto a dimensionally 
reduced twisted Eguchi-Kawai matrix model. In this formulation we 
investigate the Wilson loops and in particular the Creutz ratios.
This is an ongoing project which aims at non-perturbative predictions
for the photon, which can be confronted with phenomenology in order
to verify the possible existence of non-commutativity in nature.
\end{abstract}


\maketitle

\section{THE NON-COMMUTATIVE U(1) MODEL}

We investigate non-perturbatively QED on a non-commutative (NC) 
space-time \cite{Dou01}, where
a commutation relation of the form
\begin{equation}
[\hat{x}_\mu, \hat{x}_\nu]=i\Theta_{\mu \nu}
\end{equation}
is introduced. This has the effect of replacing points in regular space 
with ``Planck cells''. $\Theta_{\mu \nu}$ is a real and antisymmetric 
\emph{non-commutativity tensor}.


Here we address pure U($1$) gauge theory in a Euclidean NC space-time. 
Our main tool is the {\em star product}, which is defined as
\begin{equation}
f(x) \star g(x) = f(x) \exp \left( \frac{i}{2} 
\overleftarrow{\; \partial_\mu}\,\Theta_{\mu \nu}\,
\overrightarrow{\partial_\nu}\right) g(x) \ .
\end{equation}
This definition absorbs the ``space deformation'' into the products
of fields, thereby allowing us to treat the space-time operators 
$\hat{x}$ and functions thereof as regular space-time coordinates again, 
\begin{equation}
\hat{\phi}(\hat{x}) \, \hat{\psi}(\hat{x}) \rightarrow \phi (x) 
\star \psi (x) \ .
\end{equation}

Equipped with this tool we are able to write down the gauge action 
of the pure U($1$) theory as 
\begin{equation}
S[A]=-\frac{1}{4 g^2} \int \mathrm{d}^4 x \, 
\mathrm{Tr}\left[ F_{\mu \nu}(x) \star F_{\mu \nu}(x)  \right] ,
\end{equation}
where $F_{\mu \nu}$ is given by
\begin{equation}
\label{F_def}F_{\mu \nu}= \partial_\mu A_\nu - \partial_\nu A_\mu -i(A_\mu 
\star A_\nu - A_\nu \star A_\mu).
\end{equation}

The gauge fields in this theory do not commute, as the Yang-Mills term shows. 
Another unusual feature of this theory are complex, $\star$-gauge invariant
Wilson loops \cite{IIKK}. 
Moreover a $\Theta$-deformed  dispersion relation for the photon is 
expected \cite{Mat00}. Such a modification was observed
for the NC $\lambda \phi^4$ model \cite{phi4}.
This is due to a remarkable feature of the NC field theories, the UV/IR 
mixing of divergences \cite{UVIR}. 
That property makes the perturbative treatment beyond one loop
extremely difficult.

\section{LATTICE FORMULATION OF THE NC PLANE}
We carried out simulations involving two NC directions, as well as 
two commutative directions, which include the Euclidean time. To be
explicit, we set 
$\Theta_{1 2}=-\Theta_{2 1} = \theta,\textrm{ and }\Theta_{\mu \nu}=0
\textrm{ otherwise}$. The parameter
$\theta$ was assumed to be constant throughout the space-time.

In order to perform Monte Carlo simulations one needs to formulate 
the theory on a lattice. This is achieved by the operator relation
\cite{Amb99}
\begin{equation}
\exp \left( \frac{2 \pi i}{a} \, \hat{x}_\mu \right) 
= \hat{\mathbb{1}},\qquad \mu=1,2 \ ,
\end{equation}
which imposes a discrete non-commutative space-time.
After having obtained the lattice space-time, it needs to be 
populated by U($1$) link variables. The key to render their simulation ---
involving the non-local $\star$-product ---
feasible is the \emph{Morita equivalence} \cite{Aoki} of
%
%
NC U$(N)$ gauge theories to the 
\emph{twisted Eguchi-Kawai} (TEK) model \cite{Gon83}, with the action 
\begin{equation}
\label{TEK_action}S_{\mathrm{TEK}}[U]=-\beta\sum_{\mu \neq \nu} 
\mathcal{Z}_{\mu \nu} \mathrm{Tr} \left( U_\mu U_\nu U_\mu^\dagger 
U_\nu^\dagger \right).
\end{equation}
Here the $U_\mu$ are U($N$) matrices. The twist 
$\mathcal{Z}_{\mu \nu}=\mathcal{Z}^*_{\nu \mu}$ is given by 
$\exp \left( \frac{2 \pi i \, n_{\mu \nu}}{N} \right)$, with 
$n_{\mu \nu} \in \mathbb{Z}.$ In our mapping we chose 
$n_{21} = - n_{12} = (N+1)/2$ for some odd $N$, and $n_{\mu \nu} =0$ 
otherwise.

We discretized the commutative plane to a regular $(N-1) \times (N-1)$ 
lattice, where on each lattice site the NC plane is represented by a 
$N\times N$ TEK model ($N$ is odd).

The non-commutativity parameter $\theta$ and the lattice parameters 
$N$ and $a$ are related as \cite{Amb99}
\begin{equation}
\label{noncomm_lattice}\theta=\frac{1}{\pi} N a^2\;.
\end{equation}
In the $N \rightarrow \infty,$ $a \rightarrow 0$ limit this theory 
describes a continuum gauge theory; which theory one obtains
depends on the way this limit is taken. 
Keeping $\theta$ finite in the continuum and infinite volume
corresponds to the {\em double scaling limit}.
In the case of finite $N$ and $a$ the theory maps exactly to 
a NC U($1$) lattice theory \cite{Amb99}.

In order to be able to apply the heat bath algorithm, the 
action was linearized in $U_{\mu}$, as a generalization of
the scheme of Ref.\ \cite{Fab84}. 

\section{RESULTS}
\begin{figure}[t]
\begin{center}
\scalebox{0.6}{\includegraphics{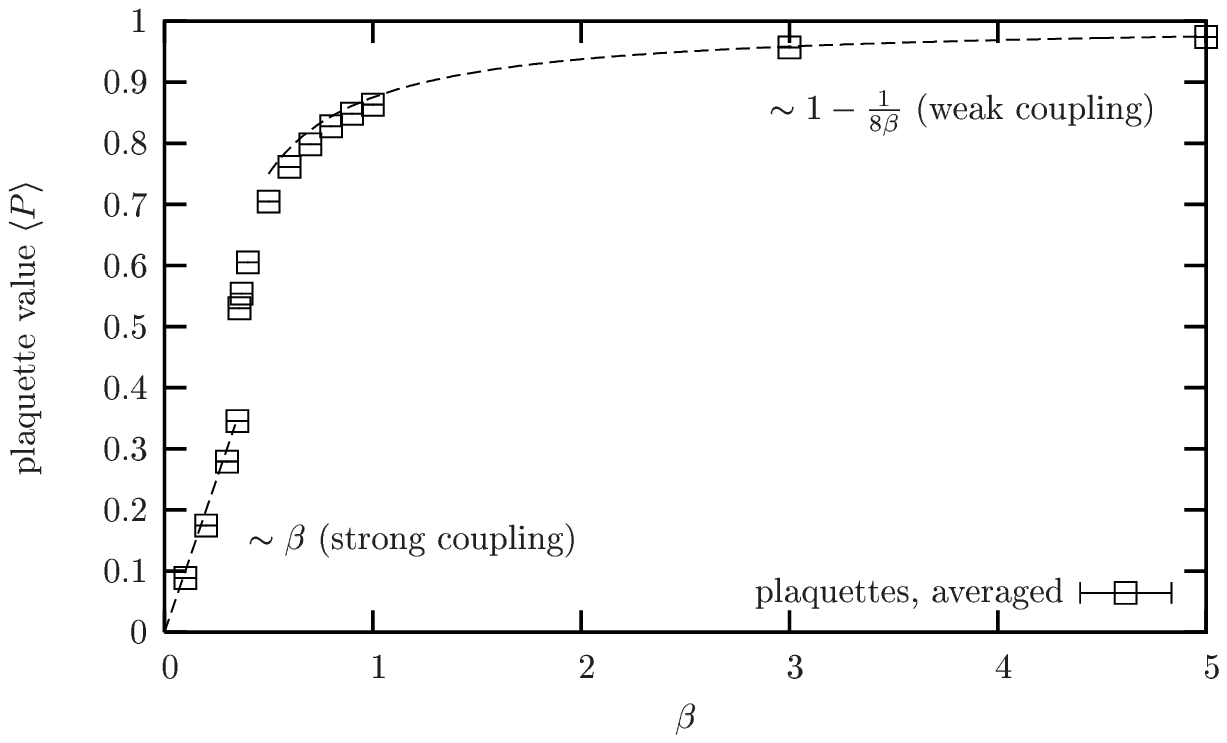}} 
\\
\scalebox{0.6}
{\includegraphics{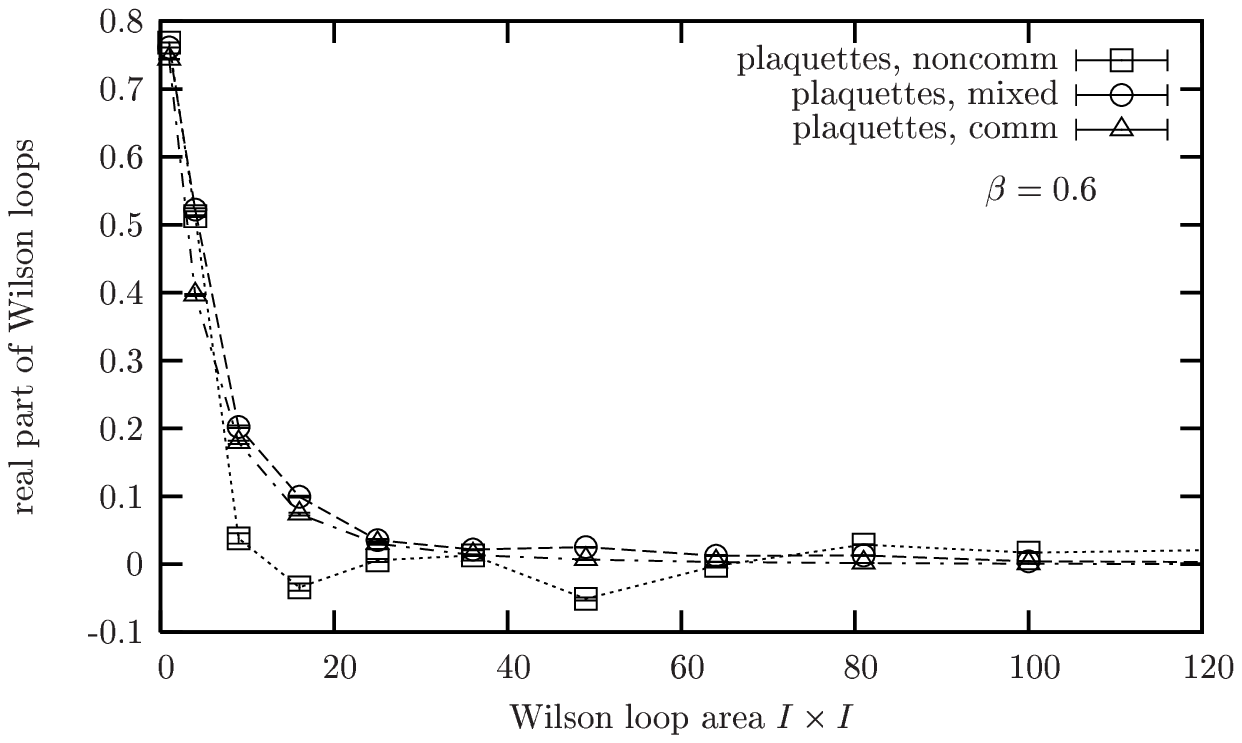}}
\vspace{-30pt}
\caption{\emph{Above: The real part of the plaquettes, averaged over all possible 
orientations. The dashed lines show the asymptotic behavior expected based on
a strong/weak coupling expansion. 
Below: The real part of the plaquettes, split according to their orientation.}}
\label{plaquettes}
\end{center}
\vspace{-1cm}
\end{figure}

In the simulation results presented here we set $N=25$.
The upper part of Fig.\ \ref{plaquettes} shows the real part of the
plaquette values at different $\beta$. In this case all six possible orientations 
of the plaquettes are averaged. The plot shows that the results match the 
anticipated asymptotic behavior \cite{Gon83}.
The phase transition occurs at around $\beta \approx 0.35$.

The decay of Wilson loops observed from differently sized squares is shown in 
the lower part of Fig.\ \ref{plaquettes}. In that plot the results are 
split according to the plaquette orientations (NC, mixed and commutative). The 
real part of the NC Wilson loops oscillates, while eventually decaying 
towards zero. This is in qualitative agreement with the observations 
in Ref.\ \cite{NCQED2d} for $d=2$. 
For the other two types we see a monotonous decay towards zero. 
Notice that these results were obtained in the weak coupling phase.
\begin{figure}[t]
\begin{center}
\scalebox{0.6}{\includegraphics{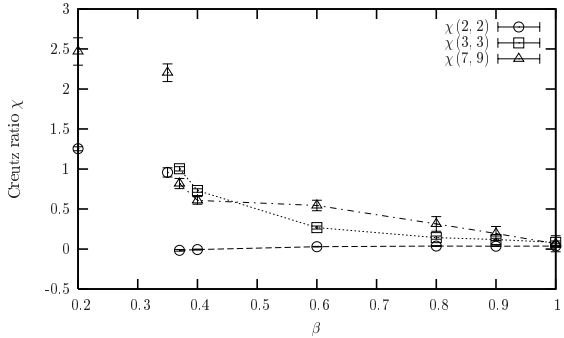}}
\vspace{-40pt}
\caption{\emph{Different size Creutz ratios $\chi$ of rectangular
Wilson loops, averaged over all planes, at various couplings.}}
\label{creutz}
\end{center}
\vspace{-1cm}
\end{figure}

We have calculated Creutz ratios $\chi$ for various rectangular Wilson loops. 
The results are shown in Fig.\ \ref{creutz}. From these we see that the string 
tension seems to approach zero as $\beta \rightarrow \infty$, which corresponds 
to the continuum limit. 
\begin{figure}
\begin{center}
\scalebox{0.6}{\includegraphics{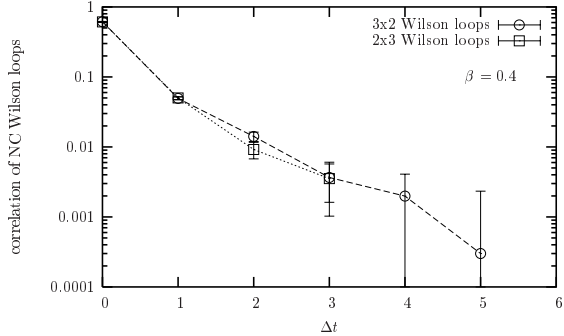}}
\vspace{-40pt}
\caption{\emph{Temporal correlation of the plaquettes lying completely 
in the NC plane.}}
\end{center}
\vspace{-1cm}
\label{noncomm_corr}
\end{figure}
\par
Fig.\ \ref{noncomm_corr} shows the correlation function of the pla\-quettes 
lying completely in the NC plane, and separated by $\Delta t$ in Euclidean time. 
These data were taken at 
$\beta=0.4,$ so we are barely in the weak coupling phase. 
The decay seems to be exponential, but more statistics are 
required to confirm this behavior.

\begin{figure}[t]
\begin{center}
\scalebox{0.6}{\includegraphics{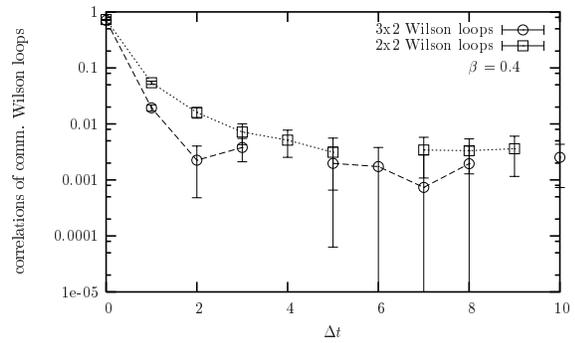}}
\vspace{-40pt}
\caption{\emph{Temporal correlation of the plaquettes lying 
completely in the commutative plane.}}
\label{comm_corr}
\end{center}
\vspace{-1cm}
\end{figure}

Fig.\ \ref{comm_corr} shows the commutative analog to the correlation 
function discussed above. However, in that case the decay does {\em not} 
seem exponential.
 

\section{SUMMARY AND CONCLUSION}
We have simulated a NC U($1$) field theory.
First observables such as the action, plaquettes, Wilson loops and Creutz 
ratios could be measured. The results for the action and plaquettes agree 
with asymptotic predictions. There seems to be a vanishing string 
tension at weak coupling in the part of the parameter space we explored. 

Further work will hopefully allow us to identify a physical scale and
to extrapolate to a NC continuum limit by means of double scaling (i.e.\
a fixed product $Na^{2}$). Thus we hope 
to glimpse at the dispersion relations for the NC photon, which might
allow us to confront the theory with nature.\\

\noindent
{\em Acknowledgement:} \ \
We would like to thank the ``Deutsche Forschungsgemeinschaft'' (DFG) for 
generous support. The computations were performed on the IBM p690 
clusters of the
``Norddeutscher Verbund f\"ur Hoch- und H\"ochstleistungsrechnen'' (HLRN).

\end{document}